%% file: setterberg-spie.tex
\documentclass[]{spie}

\usepackage{amsmath,amsfonts,amssymb}
\usepackage{graphicx}
\usepackage[colorlinks=true, allcolors=blue]{hyperref}

\title{Geant4 Modeling of a Cerium Bromide Scintillator Detector for the IMPRESS CubeSat Mission}
\input{authors.tex}

\graphicspath{{./figures/}}

\begin{document}

\maketitle
\begin{abstract}
    Solar flares are some of the most energetic events in the solar system and can be studied to investigate the physics of plasmas and stellar processes.
    One interesting aspect of solar flares is the presence of accelerated (nonthermal) particles,
        whose signatures appear in solar flare hard X-ray emissions.
    Debate has been ongoing since the early days of the space age as to how these particles are accelerated,
        and one way to probe relevant acceleration mechanisms is by investigating short-timescale (tens of milliseconds) variations in solar flare hard X-ray flux.
    The Impulsive Phase Rapid Energetic Solar Spectrometer (IMPRESS) CubeSat mission
        aims to measure these fast hard X-ray variations.
    In order to produce the best possible science data from this mission,
        we characterize the IMPRESS scintillator detectors using Geant4 Monte Carlo models.
    We show that the Geant4 Monte Carlo detector model is consistent with an analytical model.
    We find that Geant4 simulations of X-ray and optical interactions explain observed features in
        experimental data, but do not completely account for our measured energy resolution.
    We further show that nonuniform light collection leads to double-peak behavior at the 662 keV \textsuperscript{137}Cs photopeak
        and can be corrected in Geant4 models and likely in the lab.
\end{abstract}


\section{Introduction}
\label{sec:intro}
A solar flare is an impulsive release of electromagnetic energy driven by magnetic reconnection in the corona.
It has been estimated that up to 50\% of energy converted during a solar flare's magnetic reconnection process
    contributes to particle acceleration \cite{lin-accel}.
However, debate has been ongoing since the early days of the space age as to how these particles are accelerated.
Understanding the mechanisms that accelerate particles is important 
    to understand stellar processes
    and space weather.

Hard X-rays (HXRs) are produced by accelerated electrons and hot (Megakelvin)
    temperatures during solar flares via bremsstrahlung and atomic transitions.
One way to probe particle acceleration mechanisms is by investigating short-timescale (tens of milliseconds) variations in solar HXR observations,
    also known as HXR spikes.
If we can constrain the timescales of these spikes,
    then we can constrain the acceleration timescales of the electrons that produced them.

The IMPRESS CubeSat currently under development at the University of Minnesota SmallSat Research Laboratory (UMN SSRL)
in collaboration with
    Montana State University (MSU),
    the University of California, Santa Cruz (UCSC),
    and Southwest Research Institute (SwRI)
    aims to measure HXR spikes on timescales of tens of milliseconds.
It features four custom cerium bromide scintillator detectors investigating a hard X-ray energy range of 10-300 keV
    read out by Bridgeport Instruments electronics,
    and one AmpTek X-123 FastSDD silicon drift detector sensitive to soft X-ray (SXR) energies of 1-12 keV.
We will use the scintillators to investigate flare nonthermal bremsstrahlung emission
    and the X-123 to constrain thermal bremsstrahlung emission.
IMPRESS will record solar flare HXR emission at a frame rate/cadence of 32 Hz in order to measure fast
    HXR spikes and constrain acceleration mechanisms,
    and SXR emission at a 1 Hz frame rate.

In order to maximize the science output of IMPRESS, we must well-characterize its detectors.
We investigate the CubeSat's scintillator detectors using Geant4 Monte Carlo models.
We show these models are consistent with an analytical model first developed by Knuth (2017) \cite{knuth-model}.
Scintillators convert X-rays and other high-energy particles into optical photons,
    so using Geant4 optical physics, we simulate these scintillated
    optical photons to better understand detector behavior.
We verify the effect of scintillator crystal optical finish on light collection.
We further quantify the effect of interaction position within the crystal on light collection,
    and we explain some observed experimental artifacts using this information.

\section{Scientific background}
\subsection{Particle acceleration mechanisms}
There are various mechanisms that can accelerate particles during solar flares.
One mechanism is acceleration via electric fields aligned with the magnetic field present in a solar flare plasma.
Electric fields like this can be produced during the magnetic reconnection process \cite{parker-tb},
    and they can accelerate electrons to energies of tens of keV \cite{aschwanden-tb}
    which can lead to nonthermal bremsstrahlung emission.
Electric field acceleration is often classified by the Dreicer field strength $E_D$.
For $E > E_D$, a Maxwellian population of accelerated electrons can be runaway accelerated \cite{phys-sol-flare}.
Other acceleration mechanisms involve moving magnetic mirrors,
    such as first- and second-order Fermi acceleration\cite{fermi-og, jones-similar}.
These mechanisms can accelerate particles to keV energies which result in HXR bremsstrahlung emission\cite{aschwanden-tb}.
There are many other proposed mechanisms that can accelerate particles to keV energies\cite{zharkova-accel},
    such as those related to Alfv{\'e}n wave energy transport\cite{Fletcher-hudson-alfven} or
    gyrosynchrotron resonance\cite{aschwanden-tb}.

\subsection{Previous HXR spike investigations}

Kiplinger et al. (1983) used Solar Maximum Mission (SMM) HXR data
    to identify HXR spikes on timescales $\tau \sim 40$ ms
    for energies $E \in [30, 500]$ keV.
The authors used this spike data to support a strong argument for the presence of nonthermal
    electron energization during solar flares \cite{kiplinger-early}.

Further observational work has been performed on solar HXR data in an attempt to constrain acceleration timescales and mechanisms.
Aschwanden (1996) investigated delays between low- and high-energy HXR pulsations and concluded that
    the delays could be well-accounted for by time-of-flight differences between electrons of different kinetic energy,
    assuming that the electron acceleration is localized to the top of a coronal flare loop and not near the chromosphere \cite{aschwanden-tof}.
This further implies that the actual mechanisms accelerating particles must do so on timescales much faster than the delays between spikes.

Qiu et al. (2012) performed a deconvolution of data from the Reuven Ramaty High Energy Solar Spectroscopic Imager (RHESSI)
    to identify spikes of timescales $\tau < 1$ s \cite{qiu-deconv}.
This deconvolution was necessary because RHESSI used rotation (half-period $T \approx 2$ s) to time-modulate its signal for imaging.
The authors also found energy-dependent lags with higher-energy spikes being recorded first,
    consistent with the time-of-flight kinetic energy picture painted by Aschwanden (1996) \cite{aschwanden-tof}.
However, in a statistical study of RHESSI data by Cheng et al. (2012),
    the time lag between some low- and high-energy spikes was found to be inconsistent with Aschwanden's time-of-flight picture 
     \cite{cheng-statistical-rhessi}.

Altyntsev et al. (2019) investigated a 2011 flare using data from the Fermi Gamma-ray Burst Monitor (Fermi/GBM) and other instruments.
    These authors identified spikes on timescales $\tau < 1$ s,
    and further localized the acceleration site to near the top of a coronal loop using microwave measurements.
Their results are again consistent with Aschwanden's time-of-flight idea \cite{altyntsev-spikes}.
Glesener \& Fleishman (2018) inferred short ($< 1$ s) electron acceleration timescales in Konus-Wind data,
    and combined these observations with modeling to understand how accelerated electrons escape the Sun \cite{glesener-fleishman-2018}.
Finally, Knuth \& Glesener (2020) identified fast HXR spikes 
    using data from Fermi/GBM in the same 2011 flare as analyzed by Altyntsev et al. and one other flare.
They found a roughly exponential or power law distribution of spike durations with the shortest being $\sim 0.2$ s,
    near the GBM sensitivity limit given the spike-selection criteria used \cite{kg-gbm-spikes}.

All of these studies conclude that the HXR spikes are due to nonthermal particles.
We know that HXR spikes occur on timescales of tens of milliseconds to a few seconds;
    however, a lower limit on spike duration has not yet been found.
Thus, a solar-dedicated instrument capable measuring fast HXR spikes with decent energy resolution
    (at least 80\% FWHM at 20 keV \cite{knuth-thesis}) is highly desirable.

\section{The IMPRESS mission}
\label{sec:imp-mis}

\subsection{Mission overview}
\begin{figure}
    \centering
        \includegraphics[width=0.8\textwidth]{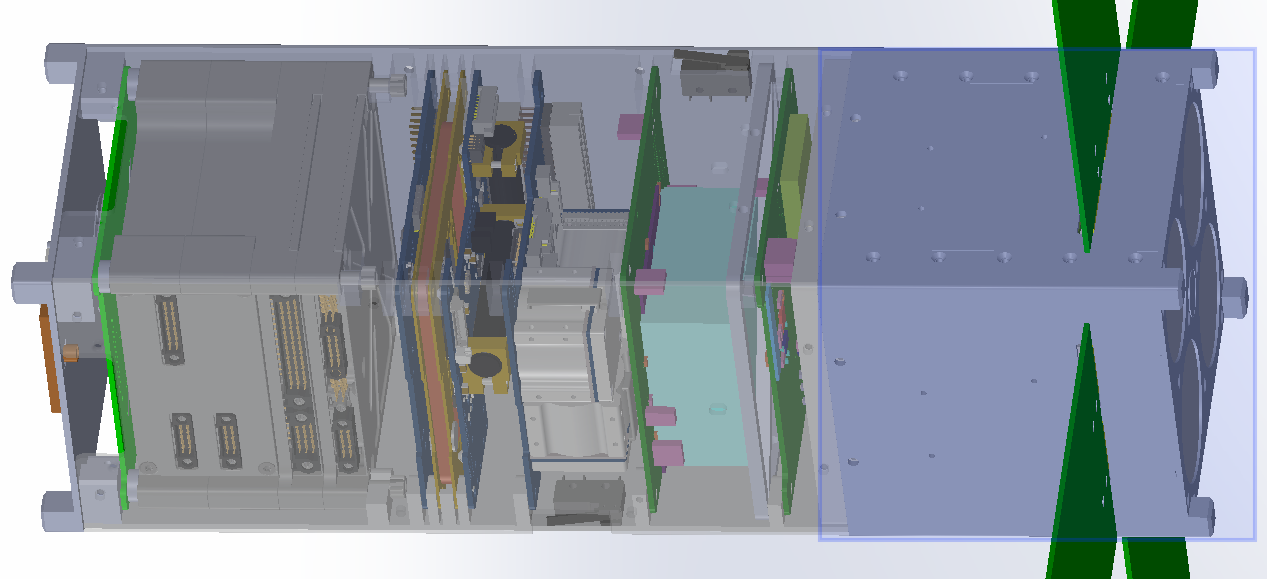}
    \caption[]{
        A cutaway view of the IMPRESS CubeSat.
        The instrument is boxed, blue, on the right;
            the four circles are the HXR detector apertures,
            and the small circle in the middle is the SXR detector aperture.
        The support equipment in the middle is for navigation, communication, power, and data handling.
        IMPRESS will launch in late 2023 or 2024 during the peak of solar cycle 25.
    }
    \label{fig:imp-cutaway}
\end{figure}

IMPRESS is a $10 \times 10 \times 30$ cm\textsuperscript{3} CubeSat mission
    currently under development at the UMN SSRL
    in collaboration with MSU, SwRI, and UCSC.
It is specifically designed to measure fast (tens of milliseconds) HXR variations produced by solar flares,
    and we are working to understand its hard X-ray detectors via Monte Carlo simulations.
Details of the IMPRESS mission and its goals are described in-depth in Ref~\citenum{knuth-thesis}.
Its mission goals in the context of measuring fast HXR spikes produced by solar flares are summarized:
\begin{enumerate}
    \item
        \textbf{Mission Objective 1:} Measure short-timescale HXR spikes with a time resolution of $\sim 30$ ms.
        Previous studies using Fermi/GBM and SMM data found HXR spikes on timescales of the order of tens of milliseconds,
            and IMPRESS will further probe these short spike timescales to constrain particle acceleration mechanisms.
    \item
        \textbf{Mission Objective 2:} Demonstrate a HXR instrument capable of measuring flares spanning several orders of
            magnitude in intensity with no moving attenuators.
        The instrument aboard IMPRESS will have four separate detectors,
            each equipped with its own non-moving (static) attenuator
            to mitigate X-ray co-incidence (pileup) events.
        Each attenuator is optimized for a particular solar flare intensity.
    \item
        \textbf{Mission Objective 3:} Investigate HXR directivity by co-observing with Solar Orbiter (SO).
        The Spectrometer/Telescope for Imaging X-rays (STIX) instrument aboard Solar Orbiter is a modern HXR instrument that can measure
            HXR ($E \in [4, 150]$ keV) variations down to 0.5 s \cite{stix-instr}. 
        IMPRESS will co-observe with SO/STIX in an attempt to probe the directivity (angular distribution) of bremsstrahlung hard X-rays,
            which can be an indicator of electron beaming behavior expected from coronal loop acceleration \cite{rhessi-elec-props}.
\end{enumerate}


\subsection{The Hard and Fast X-ray Spectrometer (HaFX)}
\label{ssec:hafx}
\begin{figure}
    \centering
        \includegraphics[width=0.8\textwidth]{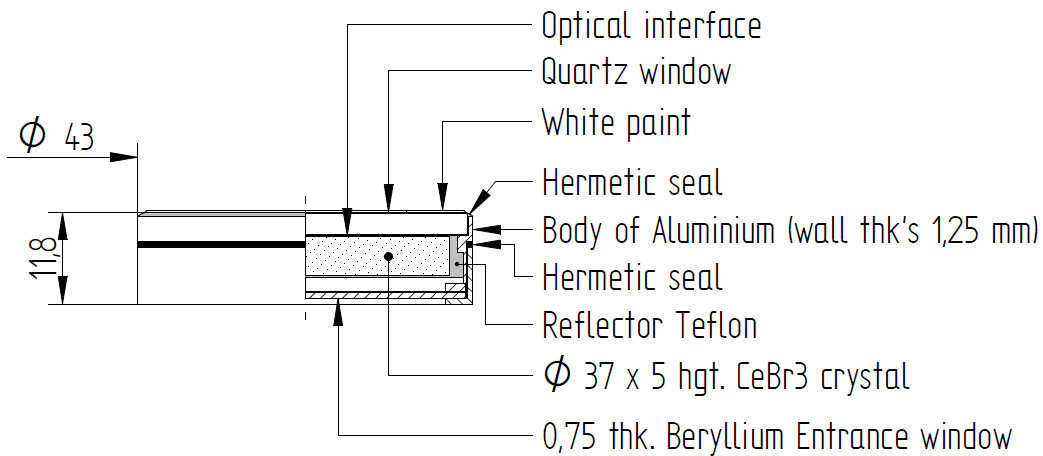}
    \caption[]{
        Schematic of our detector crystal and enclosure from Advatech (all units are in mm).
        Note the beryllium entrance window and Teflon covering the crystal.
        The detectors on IMPRESS will also have aluminum attenuator windows
            to permit measurement of different flare intensities.\\
    }
    \label{fig:schematic}
\end{figure}

HaFX is the science payload aboard IMPRESS.
It is being built by Montana State University and occupies $\approx 1000$ cm\textsuperscript{3} of the IMPRESS spacecraft (see \autoref{fig:imp-cutaway}).
It features four custom cerium bromide scintillators coupled to a $4 \times 4$
    silicon photomultiplier (SiPM) array
    and Bridgeport Instruments SiPM-3000 pulse-processing electronics.
The scintillator crystals are 37 mm in diameter and 5 mm thick,
    as detailed in the mechanical diagram in \autoref{fig:schematic}.
Each scintillator is equipped with a different-thickness aluminum attenuator window
    in order to restrict the total number of photons incident on the active volume,
    permitting measurement of several orders of magnitude of flare X-ray intensities
        (C1 through X1 Geostationary Operational Environmental Satellite [GOES] class flares).
The readout electronics will bin incident photons into an energy histogram at a 32 Hz cadence (31.25 ms time bins)
    and investigate an energy range of $E \in \left[\sim15, 300\right]$ keV.
In addition to the four scintillator detectors,
    an AmpTek X-123 FAST SDD silicon drift X-ray detector will be employed to measure thermal bremsstrahlung emission below 15 keV.

\section{Detector modeling}
\label{sec:modeling}

\begin{figure}
    \centering
        \includegraphics[height=8cm]{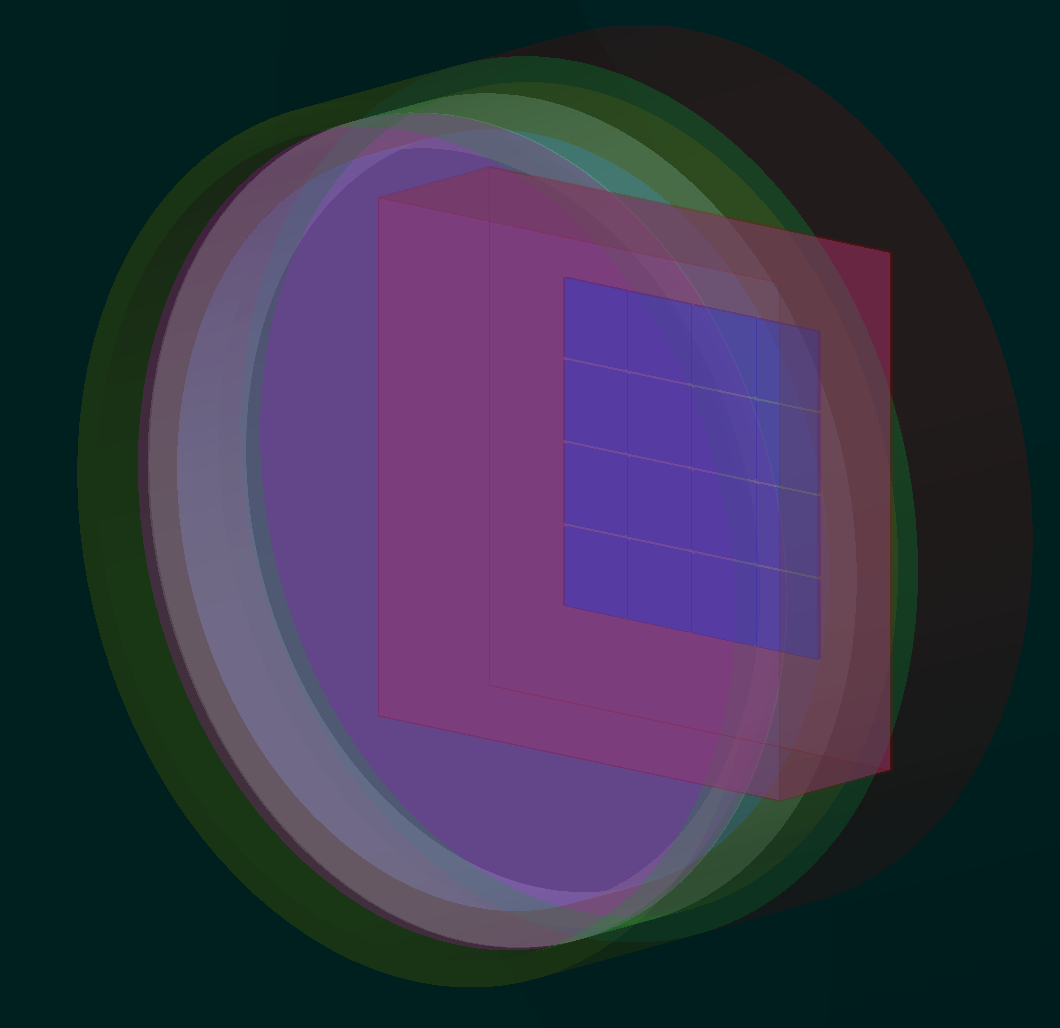}
        \includegraphics[height=8cm]{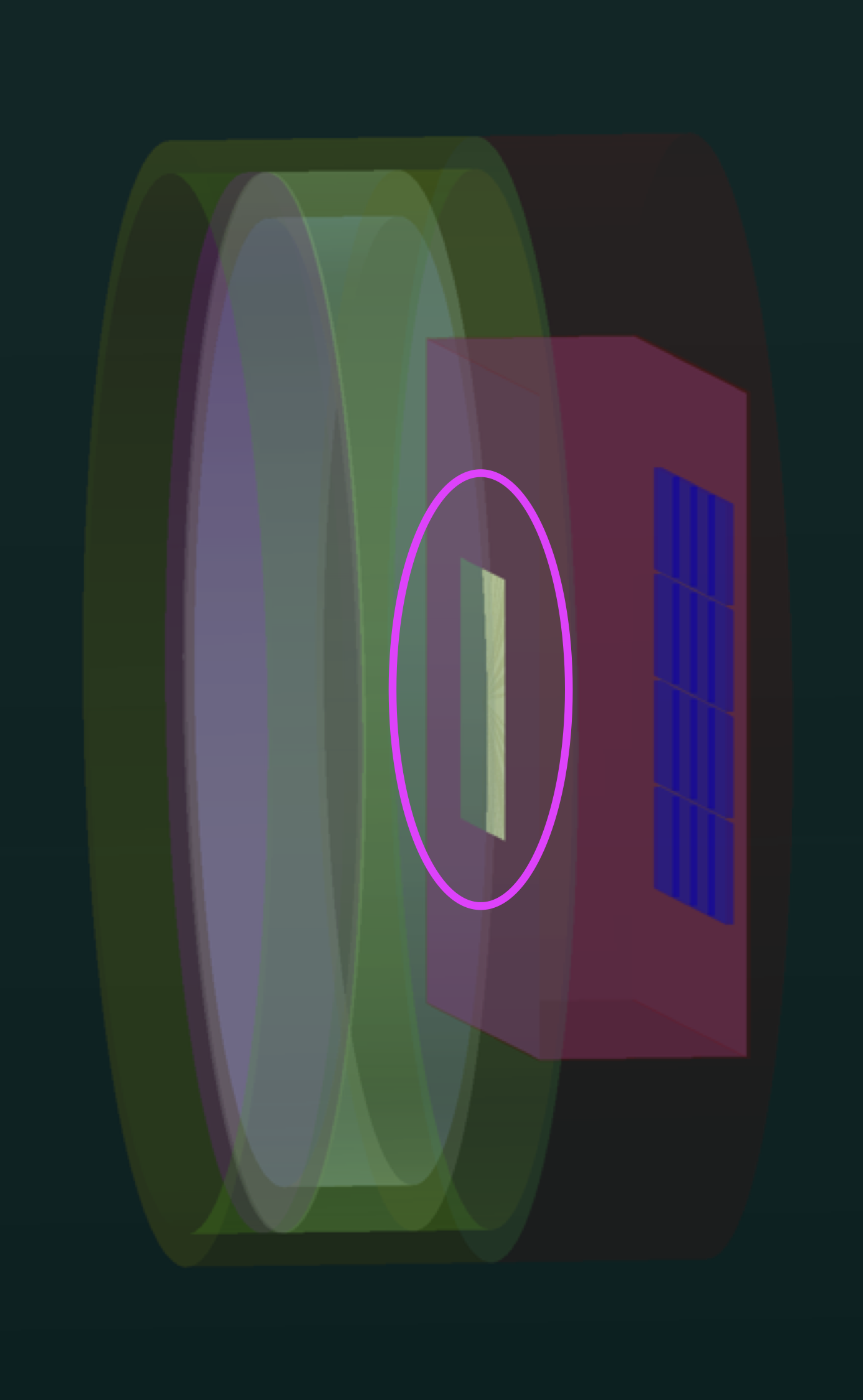}
    \caption[]{
        \textbf{Left:} The detector model in Geant4.
        The geometry from \autoref{fig:schematic} is captured in the model:
            the green sheath is the aluminum body, the pink cylinder is the beryllium window,
            the white cylindrical shell is the Teflon, the inner cylinder is the cerium bromide crystal.
            The pink light guide coupled to the crystal enclosure and the blue SiPM array are also visible.
        Components are falsely colored as a visual aid.

        \textbf{Right:} This same model with a reflective white ``shim'' inserted between the 
            detector enclosure and light guide,
            circled in magenta.
    }
    \label{fig:detector-model}
\end{figure}

Knuth (2017) developed an analytical scintillator model to be used on a similar CubeSat mission.
An adaptation of this model was employed to set IMPRESS mission requirements,
    such as aluminum attenuator window thicknesses and cerium bromide crystal dimensions.
The analytical model is an excellent starting point to understand the basics of our detector system,
    but to understand the details,
    such as the origin of our large experimentally measured energy resolution,
    or higher-order effects such as Compton scattering,
    a more complex simulation framework is necessary.
For this purpose, we have built a detector simulation using Geant4.
Geant4 is a software framework that accurately simulates the passage of particles through matter by propagating particles (in the quantum sense) one-by-one.
It has applications in high-energy particle physics, medical physics, and space physics, to name a few areas \cite{g4-validate}.
It permits accurate simulation of physical processes such as radioactive decay and all of the high-energy photon interactions
    described in e.g. Ref. \citenum{nist-db}.
It can also simulate scintillation,
    but this requires setting some empirical material parameters such as
    the photon yield (60 optical photons/keV for cerium bromide \cite{quarati-cebr3}),
    scintillation time constants,
    indices of refraction,
    and optical surface finishes.

We use Geant4 to explore the energy and scintillation properties of our detector in an attempt to understand
    and characterize its behavior.
We also use Geant4 to investigate our anomalously large experimentally-measured energy resolution.
We defined the detector geometry using information from our collaborators at MSU and the mechanical diagram in \autoref{fig:schematic}.
A screenshot of this single-detector model can be seen in \autoref{fig:detector-model}.
Eventually, the whole IMPRESS CubeSat will be incorporated into the Geant4 model so that we can account
    for measurement effects such as X-rays scattering off of the aluminum satellite enclosure into our detectors,
    permitting accurate reconstruction of the physical incident photon flux from measured counts;
    these effects will be incorporated into the instrument response matrix.

\subsection{Experimental backdrop: motivation to use Geant4}
\label{ssec:exp-backdrop}
\begin{figure}[ht]
    \centering
        \includegraphics[width=\textwidth]{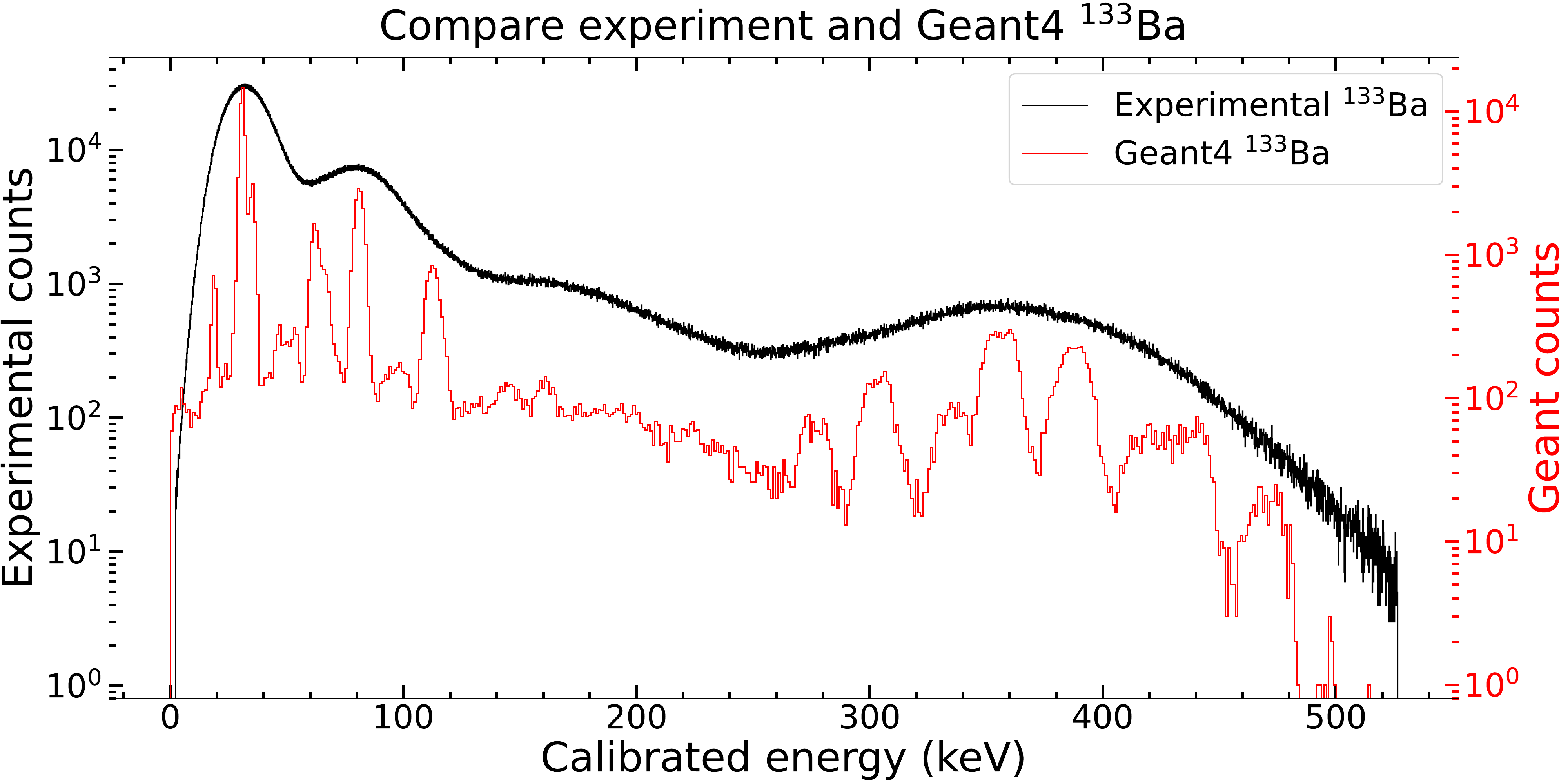}
    \caption[]{
        Comparison of Geant4 with real experimental data.
        The crystal edge and Sun-pointing face are rough-finished in this simulation.
        Both spectra have been calibrated using a linear gain calibration to convert
            detector bin (experiment) or photon counts (Geant4) to energy.
        Qualitatively the spectra have similar shapes, but Geant4 clearly has better resolution.
        Any optical photon that hits one of the SiPMs in Geant4 counts as a detection,
            so there are intrinsic resolutions that are not accounted for,
            such as electronic noise.
        The FWHM of the experimental $\sim 31$ keV peak is about 78\%,
            compared to the 8\% of Geant4.
    }
    \label{fig:g4-exp-ba133}
\end{figure}

One critical issue that the IMPRESS HXR detectors have been experiencing is anomalously large energy resolution.
The resolution of a given spectral line is usually defined as follows,
\[R = \frac{2\sqrt{2\ln2} \sigma}{E_\mu}.\]
A Gaussian profile is fit to energy-calibrated detector counts from a spectral line:
    $E_\mu$ corresponds to the center of the Gaussian (mean peak energy) and
    $\sigma$ is its standard deviation.
The factor $2\sqrt{2 \ln2}\sigma$ is the full-width at half-maximum (FWHM) of the Gaussian profile.
Fitting a Gaussian profile to the experimentally-measured 31 keV emission peak in \autoref{fig:g4-exp-ba133},
    we find a peak width of 24.4 keV, or a FWHM of about 78\%.
Quarati et al. (2013) report a peak width at this same energy of 6.0 keV, or a FWHM of about 19\% \cite{quarati-cebr3}.
Their resolution is about $4\times$ smaller than ours.
We investigate the X-ray and optical physics of our detector in detail using Geant4 to understand our large energy resolution.

\subsection{Cross-validation of analytical and Geant4 models (without optical photons)}
\label{ssec:analytical}
\begin{figure}[h]
    \centering
        \includegraphics[width=\textwidth]{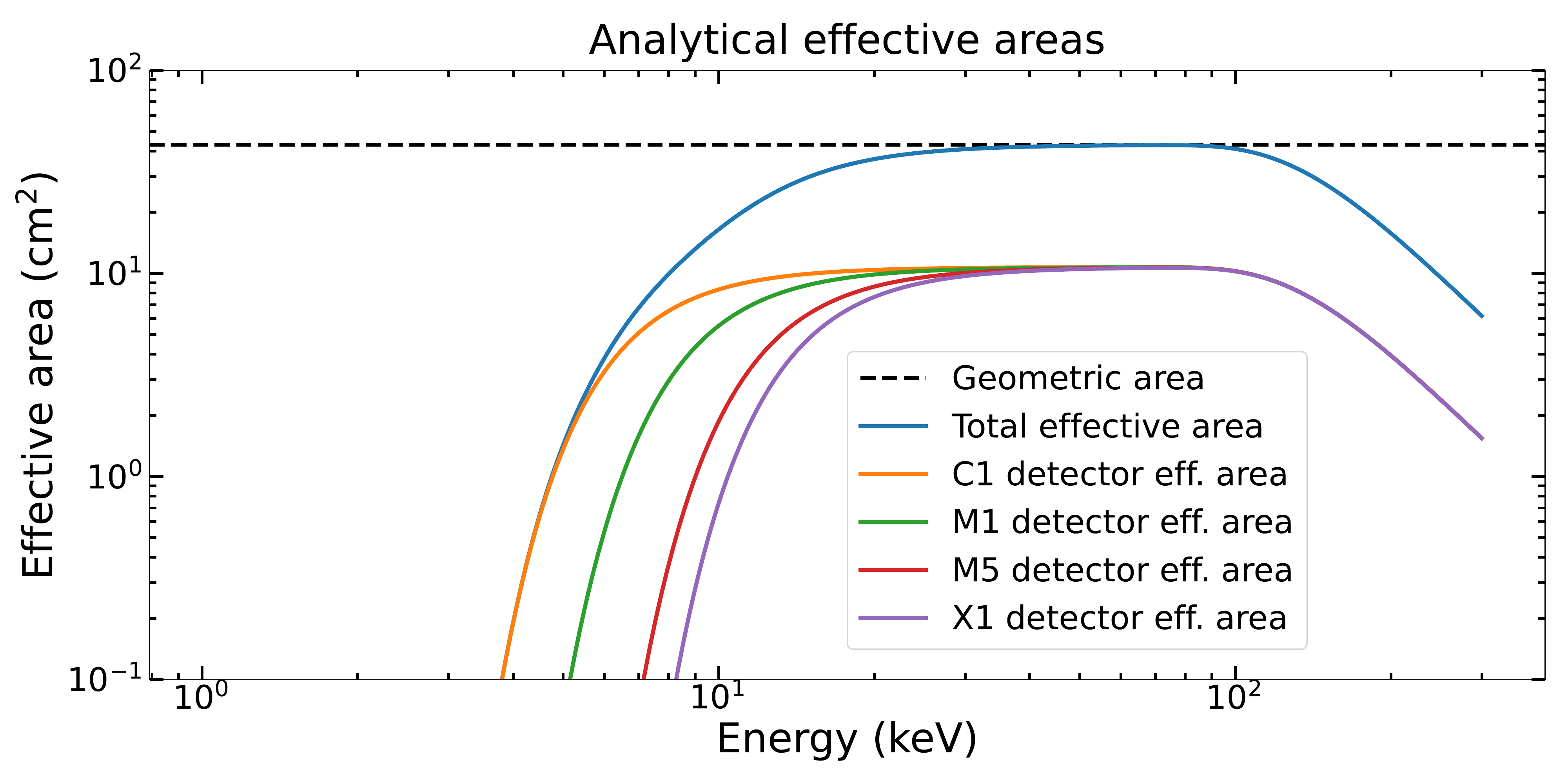}
    \caption[]{
        Effective area for each scintillator crystal channel, computed using the analytical model.
        Only Rayleigh scattering and photoelectric absorption are considered.
        Each scintillator is optimized for a different flare intensity (indicated as C1, M1, M5, and X1)
        and thus has a different-thickness aluminum attenuator,
            so each has a different effective area.
    }
    \label{fig:effa}
\end{figure}

The detectors used in IMPRESS were first modeled using an analytical model adapted from
    that described in Ref.~\citenum{knuth-model}.
This model uses mass attenuation coefficients\cite{nist-db} to quantify how an average X-ray of a given energy interacts
    with the scintillator crystal and its housing materials which are labeled in \autoref{fig:schematic}.
The mass attenuation coefficient $(\mu / \rho)$ for a particular X-ray interaction
    (photoelectric absorption, Compton scattering, etc.) is proportional to
    that interaction's cross section
    and can be used to describe interaction probability of an X-ray with a detector.
These concepts are described in detail in Ref.~\citenum{nist-db}.

Analytically, the scintillator detector is modeled as a stack of various materials as indicated in
    \autoref{fig:schematic},
    such as the aluminum attenuator window (not included in the schematic), 
    the beryllium entrance window,
    and the Teflon crystal cladding.
Each material has an associated interaction probability,
    so the total interaction probability for a particular energy is simply
    \[P_\text{tot}(E) = P_\text{Al} \cdot P_\text{Be} \cdots.\]
Only photoelectric absorption and Rayleigh scattering are considered, and scintillation photons are not considered.
Using this model, a first approximation of an effective area is obtained (\autoref{fig:effa}),
    where the effective area is just total interaction probability computed from the mass attenuation coefficients multiplied
    by the geometric area, i.e. $A_\text{effective} = P_\text{tot} \times A_\text{geometric}$.

The analytical detector model has been used to set some IMPRESS mission-critical parameters,
    such as attenuator window thicknesses,
    cerium bromide thickness,
    and geometric detector area.
It is relatively easy to use and permits fast exploration of parameter spaces.

\begin{figure}
    \centering
        \includegraphics[width=\textwidth]{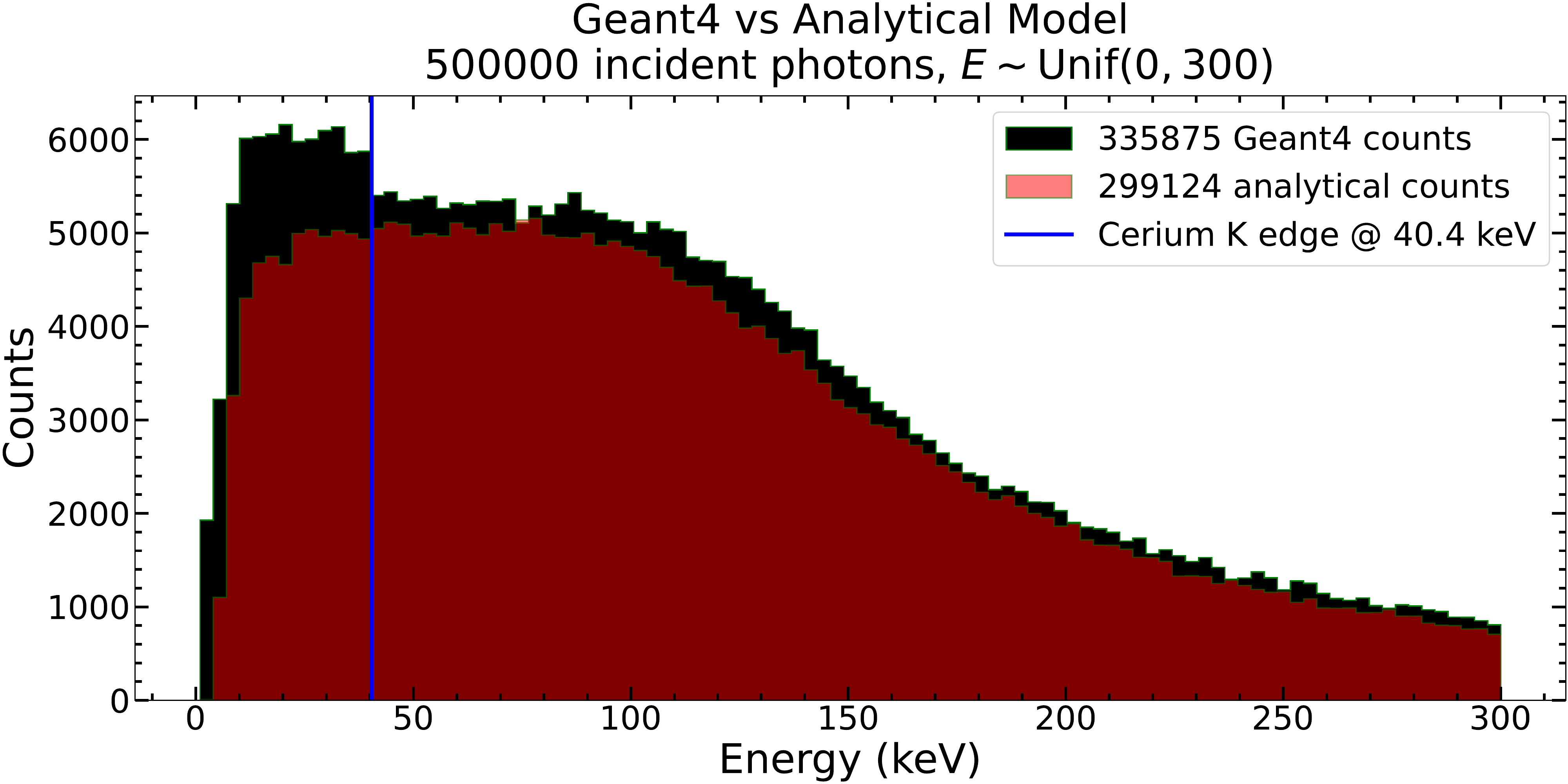}
    \caption[]{
        Comparison of the analytical and Geant4 models for a flat input spectrum, i.e. $E_{\gamma, \text{inc}} \sim \text{Unif}(0, 300)$.
        There is qualitative agreement between the two models,
            and differences are accounted for by effects included in the Geant4 model and excluded from the analytical one.
        A ``count'' here is not incorporating scintillated photons;
            a count is just an energy deposit inside the scintillator crystal.
    }
    \label{fig:ana-cmp}
\end{figure}
To cross-validate Geant4 and this analytical model,
    we illuminate both model geometries with a flat input spectrum of photons and compare the results.
\setcounter{footnote}{1}
This comparison is depicted in \autoref{fig:ana-cmp}.\footnote{
    The statistics notation $X \sim \mathrm{Unif}(a, b)$ means that the random variable $X$
        follows a uniform probability distribution between $a$ and $b$,
        i.e. its probability density function $p(x) = 1 / (b - a)$ for $x \in [a, b]$ and $p(x) = 0$ otherwise.
    }
Here the scintillation of optical photons does not take place; we are only considering energy deposition events.
There is qualitative agreement between the Geant4 and analytical model,
    but there are some distinct differences.
The Geant4 model registers more counts overall because it accounts for processes
    such as Rayleigh and Compton scattering more accurately than the analytical model.
Furthermore, there is a distinct feature at energies below the cerium K ($n = 1$) absorption edge;
    this edge is an energy at which interaction probability is sharply enhanced \cite{nist-db}.
The low-energy enhancement of counts is sharp at $E = 40.4$ keV,
    so we can conclude that
    $K$-shell ($n = 1$) transition X-ray escape must contribute heavily to the low-energy ($E < 40.4$ keV) bump;
    for a discussion of X-ray escape, see e.g. Refs~\citenum{lecoq-scint, nist-db, knoll-tb}.
These escape peaks are not incorporated into the analytical model,
    but are well-described by the Geant4 model.
Overall, the differences between the models are to be expected and are physically understood.

\subsection{Surface finish and light collection uniformity (with optical photons)}
\label{ssec:surf-finish-unif}
Prior studies have shown that cerium bromide scintillators can achieve excellent energy resolution \cite{quarati-cebr3},
    so the large energy resolution presented in \autoref{ssec:exp-backdrop} is surprising.
Knoll (2010) identifies three major contributions to scintillator detector energy resolution \cite{knoll-tb}:
1) Intrinsic scintillator resolution, due to imperfections, nonuniformities, and nonlinear effects.
2) Electronic resolution, such as system noise.
\textbf{3) Optical resolution, mainly due to how surfaces are coupled, oriented, and finished.}

MSU collaborators have found similar energy resolution using two different cerium bromide crystals from different manufacturers
    that are otherwise identical,
    so we are less interested in resolution factor (1).
MSU collaborators have also attempted to couple the scintillator crystals to photomultiplier tubes as well as SiPM arrays
    and found poor resolution in both cases,
    so we are less interested in resolution factor (2).
There is no straightforward way in the lab to probe how optical properties such as finish (rough vs polished) affect light collection,
    aside from manufacturing/modifying our own detector crystals which is not currently possible,
    so we use Geant4 to investigate resolution factor (3).

Prior studies have investigated the effect of surface finish on scintillator light collection and energy resolution,
    such as that of Ishibashi, Akiyama, and Ishii (1986) \cite{ishibashi-surface}.
Light collection is particularly important because scintillator energy resolution is a combination of various factors,
    one being counting statistics uncertainty, where $\sigma_\text{Poisson}^2 = N_\mu$,
    and $N_\mu$ is the mean number of scintillated photons detected at a particular energy.
These authors studied the scintillator bismuth germanate,
    which is a halide salt of a high-$Z$ element, similar to cerium bromide.
They found that for short and fat crystals,
    a rough surface finish is more desirable,
    while for long and skinny crystals,
    a polished finish is more desirable.
This can be understood in the sense that polished, long,
    skinny crystals act like optical fibers,
    and roughening the edges of skinny crystals causes light to scatter in such a way
    that it is less likely to reach the detection surface before being re-absorbed or otherwise lost.
Short and fat crystals on the other hand benefit from surface roughening.
This is because the diffuse scattering of the rough surface has a higher probability to scatter optical photons
    towards the SiPM array or PMT,
    which for short and fat crystals constitutes a larger average
    angular target than that of the long and skinny case.

\begin{figure}[th]
    \centering
        \includegraphics[width=\textwidth]{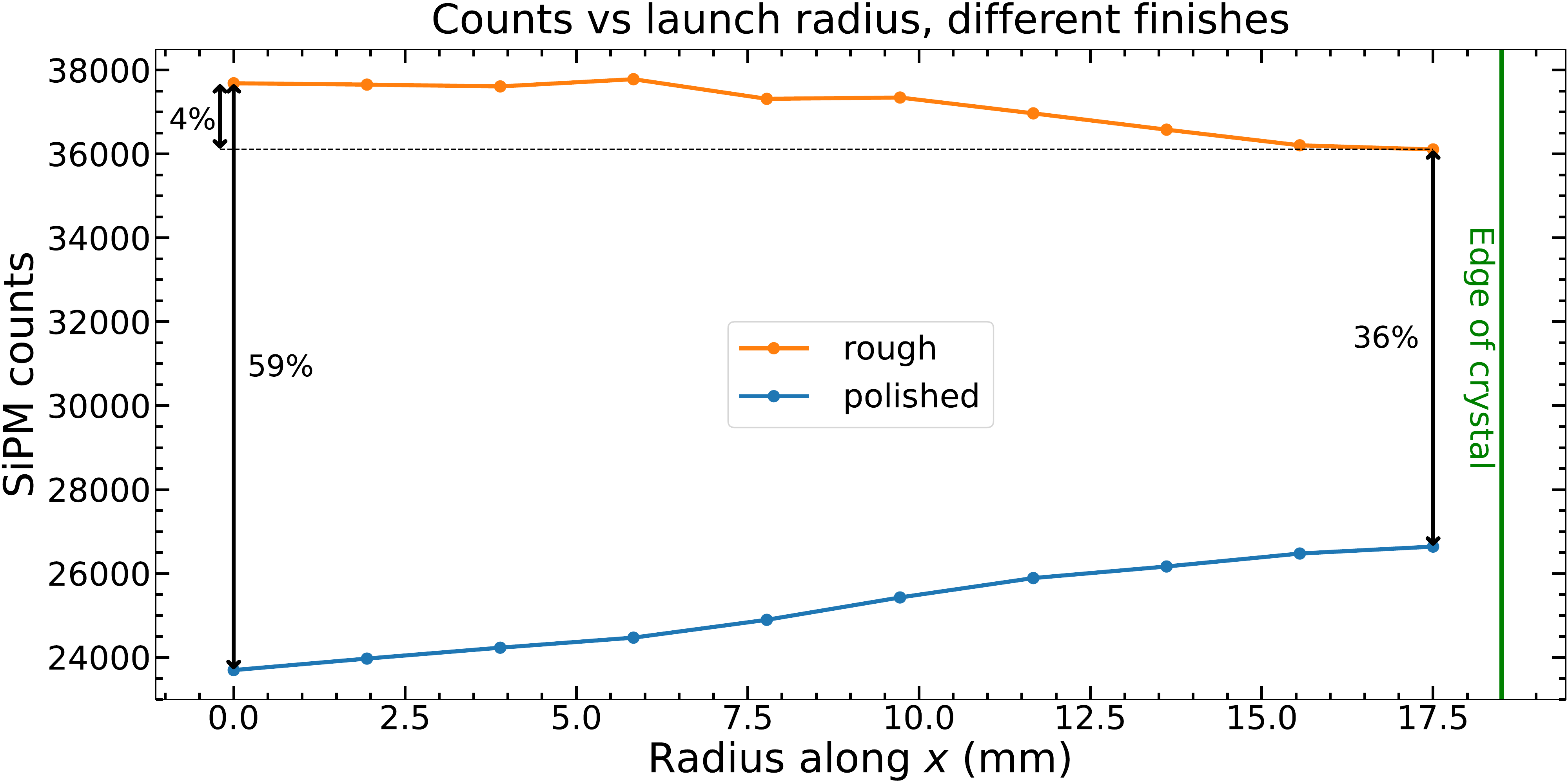}
    \caption[]{
        Light collection as a function of emission position (radius from center) for two different crystal finishes.
        50000 optical photons were isotropically emitted near the center of the crystal at fixed $z$ and $y$ while $x$ was allowed to vary;
            no X-rays were involved in this simulation.
        $z$ is aligned with the cylinder axis and $x$ and $y$ describe the plane perpendicular to this axis.
        The rough case (orange) leads to better overall light collection for our crystal geometry (37 mm diameter $\times$ 5 mm thickness).
        The $y$-axis is chosen to show the difference in collection as a function of finish and radius.
        The percent difference is taken with the polished crystal counts in the denominator,
            except for the 4\% line,
            which is taken with the counts at lowest radius in the denominator.
        The crystal surface reflection probabilities are based on experimentally-measured and tabulated values,
            described in more detail in Roncali et al. (2013) \cite{roncali-first} and (2017) \cite{roncali-update}.
    }
    \label{fig:finish-rad}
\end{figure}

The results of Ishibashi et al. (1986) are qualitatively verified in Geant4,
    as can be seen in \autoref{fig:finish-rad}.
This figure quantifies the light collection uniformity of our detector geometry
    (see \autoref{fig:schematic} for a schematic of this geometry and \autoref{fig:detector-model} for the Geant4 realization).
Across the board,
    the polished crystal leads to poorer light collection than the roughened crystal.
Light collection is actually more uniform as a function of beam position for the rough crystal than for the polished crystal:
    $\left|\Delta N\right| / N_\text{center} = 12\%$ in the polished case and $\left|\Delta N\right| / N_\text{center} = 4\%$ in the rough case,
    where $\Delta N~=~N_\text{center}~-~N_\text{edge}$.
This is again due to the fact that the rough crystal is more likely to scatter photons into the SiPM active volume than the polished one,
    and photons that undergo enough reflections are eventually absorbed into the cerium bromide crystal or other materials.

This analysis is particularly interesting because a higher photon collection rate generally improves energy resolution due to counting statistics.
Thus, a rough crystal should give us better resolution.
In the actual hardware,
    the Sun-pointing crystal face is coarsely roughened\textemdash
    the Sun-pointing face corresponds to the bottom of \autoref{fig:schematic}\textemdash
    as are the edges.
So, our physical detector is a rough crystal, but this does not solve our energy resolution issues.
Nevertheless, these results provide valuable insight for current and future instruments using scintillator detectors.

\begin{figure}[h]
    \centering
        \includegraphics[width=\textwidth]{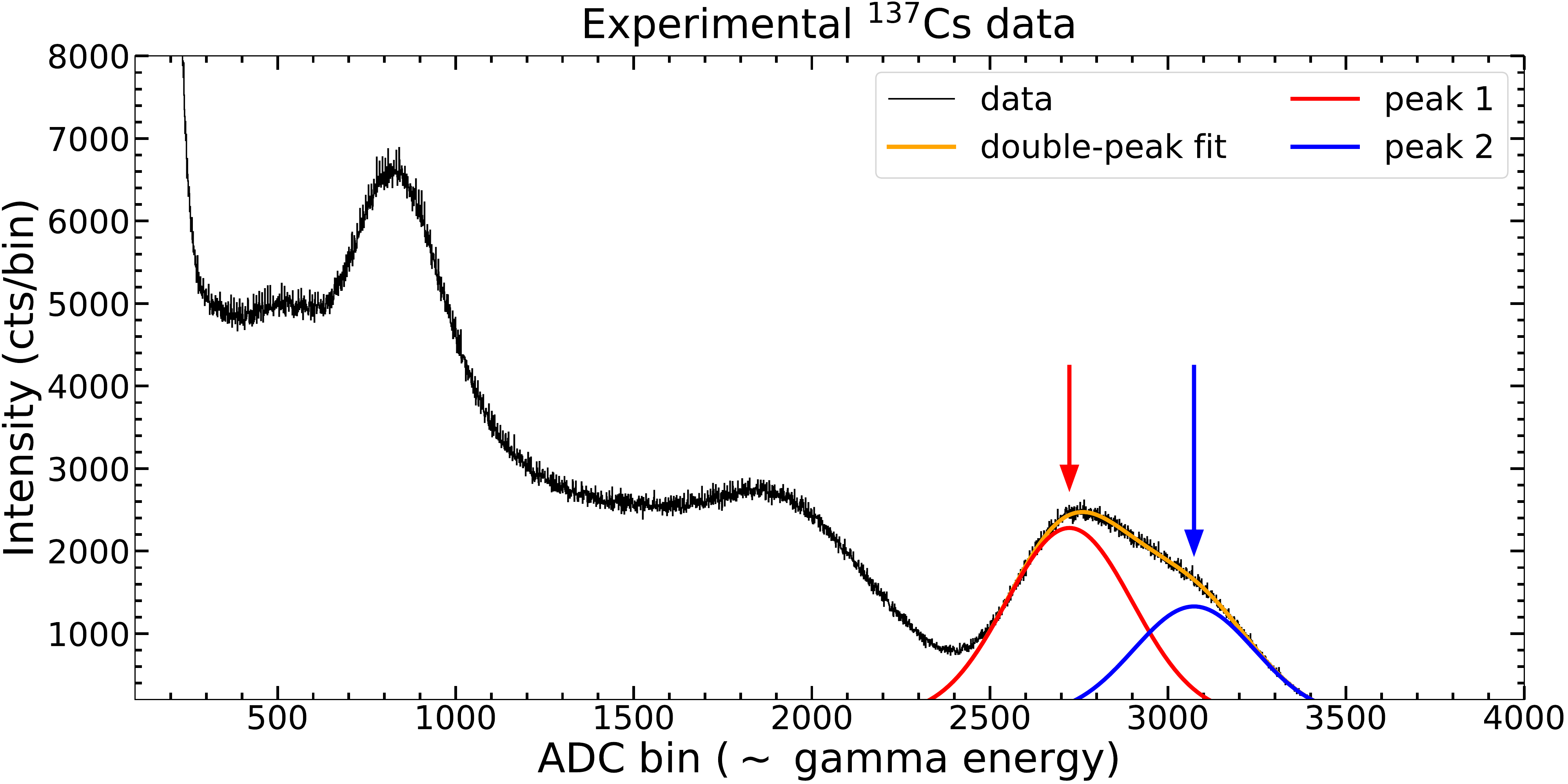}
    \caption[]{
        A 15 minute collection of ${}^{137}$Cs data measured at Montana State University by our
            37mm diameter $\times$ 5mm height cerium bromide scintillator detector.
        The red arrows call out the double-peak structure observable at the 662 keV photopeak.
        The spectrum is presented as the raw data in counts per analog-to-digital converter (ADC) bin,
            which can be mapped to incident photon energy.
        The signal corresponding to the 662 keV photopeak is fitted by a double-Gaussian function.
        Peak 1 is centered at ADC bin $\mu_1 = 2723$ and peak 2 is centered at bin $\mu_2 = 3073$,
            with $\text{FWHM}_1 = 15\%$ and $\text{FWHM}_2 = 13\%$.
        There is a $\sim 10 \%$ separation between the peaks,
            or about double the center-to-edge optical collection nonuniformity as deduced from the
            Geant4 model presented in \autoref{fig:finish-rad}.
    }
    \label{fig:cs-exp-dat}
\end{figure}

\subsection{Direct comparison of experiment and model (with optical photons)}
\label{ssec:comp-exp-model}
Geant4 can simulate radioactive decay,
    and this capability is used to directly compare an experimentally measured spectrum of a \textsuperscript{133}Ba radioisotope 
    to our Geant4 simulation.
In this simulation, optical photons are actually scintillated and detected.
The comparison results are presented in \autoref{fig:g4-exp-ba133}.
Geant4 accurately reproduces the expected spectrum of \textsuperscript{133}Ba
    and achieves a much better energy resolution than our physical detector.
The energy resolution of Geant4 is about $9\times$ better at 31 keV,
    which is somewhat expected as Geant4 ignores electronic noise and other aspects of the system,
    such as the detection efficiency of the SiPMs\textemdash every photon that hits a SiPM counts as a detection.

This being said,
    even suffering from the optical nonuniformities described in \autoref{ssec:surf-finish-unif},
    Geant4 achieves a $9\times$ better energy resolution than our physical detector.
These results reinforce the idea that \textbf{there must be other significant factors contributing to
    less-than-desirable energy resolution besides the physical optics of our detector}.
Geant4 has been validated in experiment numerous times \cite{g4-validate},
    and other researchers have achieved much better energy resolution with cerium bromide at similar energies \cite{quarati-cebr3}.

\subsection{Improving light collection uniformity via ``forced'' reflection}
\label{ssec:improve-force}
\begin{figure}[h]
    \centering
        \includegraphics[width=\textwidth]{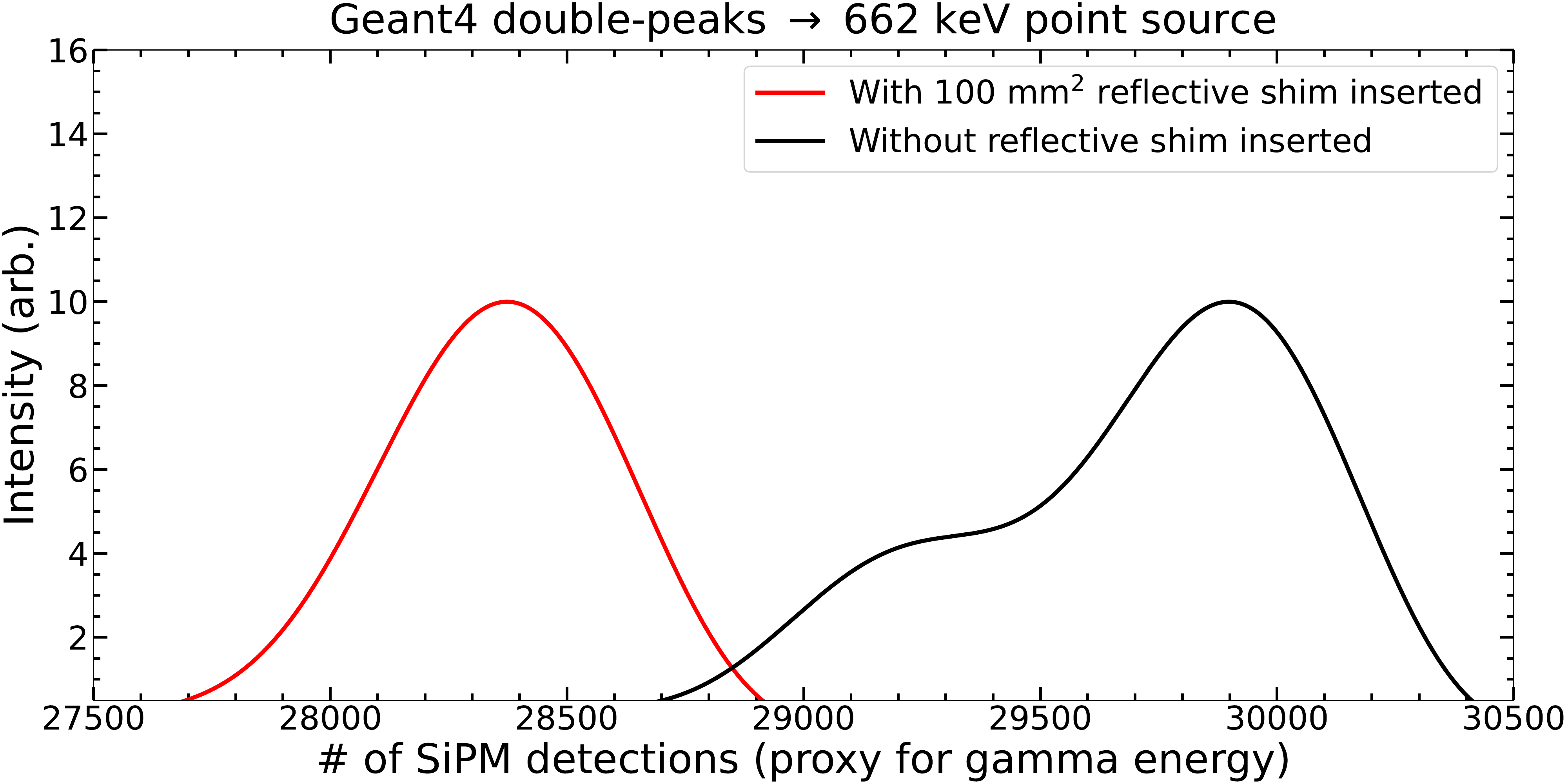}
    \caption[]{
        A monoenergetic 662 keV point source is placed $< 1$cm from the front face of the detector geometry
            (i.e. the bottom of \autoref{fig:schematic}).
        500000 gammas are emitted from the point source isotropically.
        For each gamma that interacts with the scintillator,
            we count the number of scintillated optical photons that reach the SiPM.
        The number of detections is then smoothly ``histogrammed'' using a kernel density estimator,
            and this estimator is plotted above for two cases.
        The black curve corresponds to the real-life detector geometry (\autoref{fig:detector-model}, left).
        The red curve corresponds to that same geometry with a reflective square placed between
            the detector enclosure and light guide (\autoref{fig:detector-model}, right).
        Due to nonuniform light collection and loss,
            we see a double-peak structure emerge in the black curve,
            but this is mitigated by the reflective shim employed to produce
            the red curve at the cost of collecting fewer photons on average.
    }
    \label{fig:shim-peaks}
\end{figure}

In \autoref{ssec:surf-finish-unif}, it was found that light collection varies by about 4\% from center to edge
    of our scintillator crystal.
Because the signal produced by our detectors is fundamentally a function of the number of photons counted by the SiPM array,
    this nonuniform light collection leads to experimental artifacts and
    worsens our energy resolution.
Experimentally-measured ${}^{137}$Cs data shows a double-peak structure at the 662 keV photopeak
    as depicted in \autoref{fig:cs-exp-dat}
    which we attribute to nonuniform light collection.
Although we have concluded that there must be significant factors
    other than nonuniform light collection contributing to our
    anomalous energy resolution in \autoref{ssec:comp-exp-model},
    we attempt to better understand the light collection aspect of our energy resolution
    by investigating light collection nonuniformity in the context of this 662 keV photopeak.

Light collection nonuniformity is due to crystal self re-absorption and other photon loss mechanisms;
    photons that are scintillated and directly propagate to the SiPM array undergo fewer losses on average
    than optical photons that reflect several times before reaching the SiPM array.
To improve light collection uniformity,
    we force all scintillated photons to under go some reflections
    by placing a 10mm $\times$ 10mm $\times$ 1$\mu$m reflective ``shim'' between the crystal
    enclosure and light guide as in \autoref{fig:detector-model}, right.
We chose to use a 100 mm\textsuperscript{2} shim because its
    angular size is about the same as the SiPM array from the perspective of the
    front center (bottom center of \autoref{fig:schematic})
    of our cerium bromide crystal.
Experimentally this shim could be realized as a white-paint square.
This shim effectively blocks scintillated light from directly reaching the SiPM array and
    ensures that nearly all optical photons undergo reflections.

To qualitatively compare Geant4 with experiment,
    we place an isotropically emitting 662 keV point source $< 1$cm from the front face of our detector geometry;
    500000 gammas are emitted from the point source and the detector scintillation response is recorded.
We choose this energy as it corresponds to the most intense \textsuperscript{137}Cs gamma.
In \autoref{fig:shim-peaks} we present simulation results with and without the reflective shim;
    the Geant4 results are similar to experimental ones presented in \autoref{fig:cs-exp-dat}.
With a 10mm $\times$ 10mm shim inserted between the crystal enclosure and light guide,
    the peak shape is more uniform,
    but the mean number of photons collected goes down by roughly 5\%.
So, we can resolve this double-peak behavior at the cost of worsening counting statistics.
Because we are able to reproduce this double-peak behavior in Geant4 simulations and resolve it
    by forcing scintillation photons to reflect,
    we conclude the 662 keV double-photopeak depicted in \autoref{fig:cs-exp-dat}
    is due to nonuniform light collection.
Experimental verification of these simulation results is highly desirable.

\section{Conclusions and future work}
\label{sec:conc}
Short-timescale solar flare HXR spikes have been under investigation for many years,
    but questions remain about their origins.
There is a variety of particle acceleration mechanisms that could contribute to
    observed short-timescale HXR spikes,
    and an instrument that can measure HXRs at high cadence is highly desirable.
IMPRESS will measure these spikes at a 32 Hz cadence,
    and there are outstanding questions regarding the HaFX scintillator detectors.

This paper has presented results from investigating our
    scintillator detector geometry using the Geant4 simulation framework.
We have learned about the importance of crystal finish in particular,
    and better understand effects such as secondary X-ray escape.
We have concluded that while our detector optics lead to nonuniform light collection,
    there must be other significant factors of our energy resolution as evident in \autoref{fig:g4-exp-ba133}.
Finally,
    we have shown that the nonuniform peak shape of our short and wide (5mm height $\times$ 37mm diameter) detector
    can be improved at the cost of photon counting statistics.

Analyzing real data from IMPRESS will require development of the CubeSat's instrument response matrix.
IMPRESS will provide new insight into solar flare particle acceleration,
    and Geant4 simulations will help maximize the science output of the mission.

\noindent\textit{
    PS: A basic, working, Geant4.11 example of scintillation is available on GitHub at \url{https://github.com/settwi/g4-basic-scintillation}.
    The full IMPRESS Geant4 simulation is available upon request.
}

\acknowledgments
We would like to acknowledge NSF grant number AGS1841006 for funding. 
We would also like to acknowledge further support for the UMN SmallSat Research Laboratory
    from the Minnesota Space Grant Consortium,
    the NASA CubeSat Launch Initiative,
    and the University Nanosatellite Program.
Thanks to Lukas Ley at UMN for providing the render in \autoref{fig:imp-cutaway}.
Thanks to Allan Faulkner for collecting the \textsuperscript{137}Cs and
    \textsuperscript{133}Ba experimental data.

\bibliography{setterberg-spie} 
\bibliographystyle{spiebib} 

\end{document}

%% file: authors.tex
\author[a]{William Setterberg}
\author[a]{Lindsay Glesener}
\author[a]{Demoz Gebre Egziabher}
\author[b]{John G. Sample}
\author[c]{David M. Smith}
\author[d]{Amir Caspi}
\author[b]{Allan Faulkner}

\author[a]{Lestat Clemmer}
\author[a]{Kate Hildebrandt}
\author[a]{Evan Skinner}

\author[a]{Annsley Greathouse}
\author[a]{Ty Kozic}
\author[a]{Meredith Wieber}
\author[a]{Mansour Savadogo}
\author[a]{Mel Nightingale}

\author[a, e]{Trevor Knuth}

\affil[a]{University of Minnesota, Twin Cities}
\affil[b]{Montana State University}
\affil[c]{University of California, Santa Cruz}
\affil[d]{Southwest Research Institute}
\affil[e]{NASA Goddard Space Flight Center}

\authorinfo{Contact W. Setterberg at \texttt{sette095@umn.edu}}